\documentclass{emulateapj}
\usepackage{amsmath}
\usepackage{graphicx}
\usepackage{color}

\newcommand{\beq}{\begin{equation}}
\newcommand{\eeq}{\end{equation}}

\begin{document}

\title{Mapping the Dynamics of Cold Gas around Sgr A* through 21 cm Absorption}
\author{Pierre Christian and Abraham Loeb}
\affil{Harvard Smithsonian Center for Astrophysics}
\affil{60 Garden St, Cambridge, MA.}
\email{pchristian@cfa.harvard.edu}

\begin{abstract}
The presence of a circumnuclear stellar disk around Sgr A* and megamaser systems near other black holes indicates that dense neutral disks can be found in galactic nuclei. We show that depending on their inclination angle, optical depth, and spin temperature, these disks could be observed spectroscopically through 21 cm absorption. Related spectroscopic observations of Sgr A* can determine its HI disk parameters and the possible presence of gaps in the disk. Clumps of dense gas similar to the G2 could could also be detected in 21 cm absorption against Sgr A* radio emission.
\end{abstract}

\section{Introduction}
The presence of a disk of massive young stars around the Galactic Centre \citep{StellarDisk} and various megamaser systems \citep{Miyoshi, Moran, Herrnstein, Kuo} indicates that dense neutral disks can be found in galactic nuclei. These disks could be produced by the tidal disruption of molecular clouds passing close to the central supermassive black hole \citep{YZ1}. The disk could be an outer extension of the hot accretion disk closer to the black hole, and possess a large amount of neutral hydrogen. Indeed, a significant portion of the $\rm H_2 O$ megamasers show X-ray absorption with large column densities of $ 10^{24}-10^{25} \; \rm cm^{-2}$\citep{Loeb}. Further, the discovery of the G2 cloud \citep{G2} indicates that dense clouds of self-shielded neutral hydrogen may exist around Sgr A*. 

Here we present a method to observe HI disks in the intermediate region of $10^3-10^4$ Schwarzschild radii around nuclear black holes through their 21 cm absorption. The inherent brightness of such a disk is too weak to be detectable, however the black hole's inner accretion flow emits synchrotron radiation \citep{Bower} that acts as a background source upon which the neutral systems can be seen in absorption. We examine how observations of the absorption spectrum allows one to determine the parameters of the HI absorber without the need to spatially resolve the system. 

\section{21 cm absorption profile}
\begin{figure}
\centering
\includegraphics[scale=0.2]{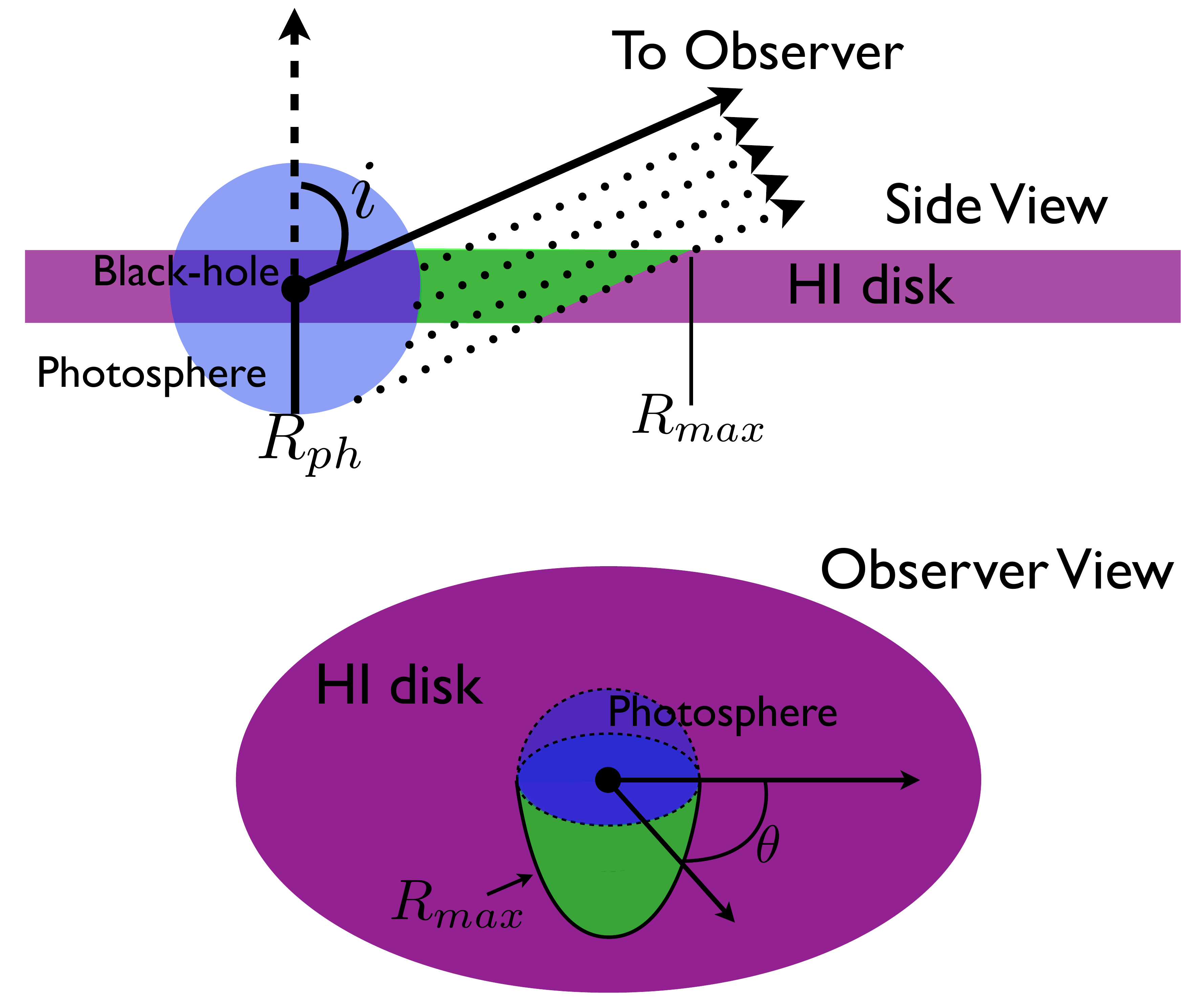}
\caption{A sketch of the geometry from the side (top image) and as seen by the observer (bottom image). The black hole is surrounded by an optically thick photosphere with a radius $R_{ph}$ and an HI disk. The solid arrow in the top image indicates the direction towards the observer. The inclination angle $i$ is the angle between the observer and the normal to the disk. The dotted arrows in the top image depict photons leaving the photosphere in the direction of the observer. In their path towards the observer, these photons encounter the HI disk and their intensity is reduced by absorption. The region of the HI disk that is illuminated by these photons is shaded green in both images. The outer radius of this region, as seen from the observer's point of view (bottom image), is referred to as $R_{max}$.}
\label{orientation}
\end{figure}

The accretion flow near Sgr A* is optically-thick to synchrotron self-absorption at a wavelength of 21 cm, and this photosphere illuminates the outer HI disk. At a distance $D$, the photosphere intensity $I_{\nu0}$ yields a flux of 
\begin{align}
F_\nu &=  \int_{\rm A} \frac{r d\theta dr}{D^2} I_{\nu0} e^{-\tau_\nu} \nonumber \\ 
&\approx \int_{\rm A} \frac{r d\theta dr}{D^2} I_{\nu0} (1 - \tau_\nu) \nonumber \; ,
\end{align}
where $I_{\nu0}$ is the photospheric specific intensity, $(r, \theta)$ are coordinates on the disk plane centered on Sgr A*, and the integral is over the area $A$ of the disk that is illuminated by Sgr A*'s photosphere. The frequency dependent optical depth is given by \citep{Loeb},
\begin{equation} \label{eq:taunu}
\tau_\nu (r, \theta)= \frac{3}{32 \pi} \frac{h^3 c^2 A_{21}}{E^2} \frac{N_H}{k T_s} \nu \phi\left[\nu - \frac{1}{c}  \sqrt{\frac{G M}{r}} \sin i \cos \theta \right] \;, 
\end{equation}
where $M$ is the black hole mass, $i$ the disk inclination defined as the angle between the observer and the normal to the disk (refer to Figure \ref{orientation}), $A_{21} = 2.85 \times 10^{-25} \; \rm s^{-1}$ is the Einstein coefficient of the 21 cm transition, $E/k= 0.068 \; \rm K$ is the 21 cm transition energy, $N_H$ is the HI column density in the disk, $\nu$ is the frequency, and $T_S$ is the gas spin temperature. For our calculations, the line function $\phi$ is a gaussian, 
\begin{align}
& \phi\left[ \nu - \frac{1}{c}\sqrt{\frac{G M}{r}} \sin i \cos \theta \;  \right]  \equiv 
\\ & \; \; \;  \frac{1}{ \sqrt{2\pi} \sigma}  \exp{- \left\{ \frac{\left[ \nu - \nu_{21} \left( 1 - \frac{1}{c}\sqrt{\frac{G M}{r}} \sin i \cos \theta \right)  \right]^2 } { 2 \sigma^2}   \right\} } \; ,
\end{align}
with  $\sigma \ll \nu_{21} \equiv  1.4 \times 10^9 \; \rm{Hz}$. For a black hole with a photosphere radius $R_{ph}$, the area $A$ can be evaluated geometrically. Assuming that the disk is larger than the photosphere, the amount of disk that is illuminated by the photosphere is the half-circular area spanned by the photosphere projected onto the disk. Only half a circle is required because the other half the disk is positioned behind the photosphere. For a disk with an inclination $i$, the projection gives
\beq
F_{\nu} = \int_{\theta=0}^{\theta=\pi}  \int_{r=R_{ph}}^{r=R_{max}} \frac{r d\theta dr}{D^2} I_{\nu_0} (1 - \tau_\nu) \; ,
\eeq
where 
\beq
R_{max} \equiv R_{ph} \sqrt{\sin^2 \theta \tan^2 i + 1 } \; ,
\eeq
is the outer radius of the region in the disk that participates in absorption (see Figure \ref{orientation}). The absolute value of the term proportional to $\tau_\nu$ can be written as
\begin{align}
\delta F_\nu &\equiv  \int_{\theta=0}^{\theta=\pi}  \int_{r=R_{ph}}^{r=R_{max} } \frac{r d\theta dr}{D^2} I_{\nu_0} \tau_\nu \nonumber  \\
&=  \int_{\theta=0}^{\theta=\pi}  \int_{r=R_{ph}}^{r=R_{max}} \frac{r d\theta dr}{D^2} I_{\nu_0} \nonumber \\
&\;\;\;\; \times \left[\frac{3}{32 \pi} \frac{h^3 c^2 A_{21}}{E^2} \frac{N_H}{k T_s} \nu  \phi\left( \nu - \frac{1}{c}\sqrt{\frac{G M}{r} \sin i \cos \theta} \;  \right) \right] \; . \label{eq:master}
\end{align}
In general, $N_H$ and $T_S$ may depend on the radial coordinate $r$. In addition, $I_{\nu_0}$ also depends on both $r$ and $\theta$ due to limb darkening. 
\subsection{Homogeneous disk with no limb darkening}
For simplicity, we focus our attention to a homogeneous disk where $N_H$, $T_S$, and $I_{\nu_0}$ are constants. From equation (\ref{eq:master}),
\begin{align} \label{eq:form}
\delta F_\nu &= \frac{3}{32 \pi} \frac{h^3 c^2 A_{21}}{E^2} \frac{N_H}{k T_s} \nu  \left( \frac{I_{\nu_0}}{D^2} \right)\int_{\theta=0}^{\theta=\pi}  d\theta \nonumber
\\ &\;\;\;\;\times \int_{r=R_{ph}}^{r=R_{max}} dr \; r \phi\left( \nu - \frac{1}{c}\sqrt{\frac{G M}{r} \sin i \cos \theta} \;  \right) \; .
\end{align}

\begin{figure}
\centering
\includegraphics[scale=1]{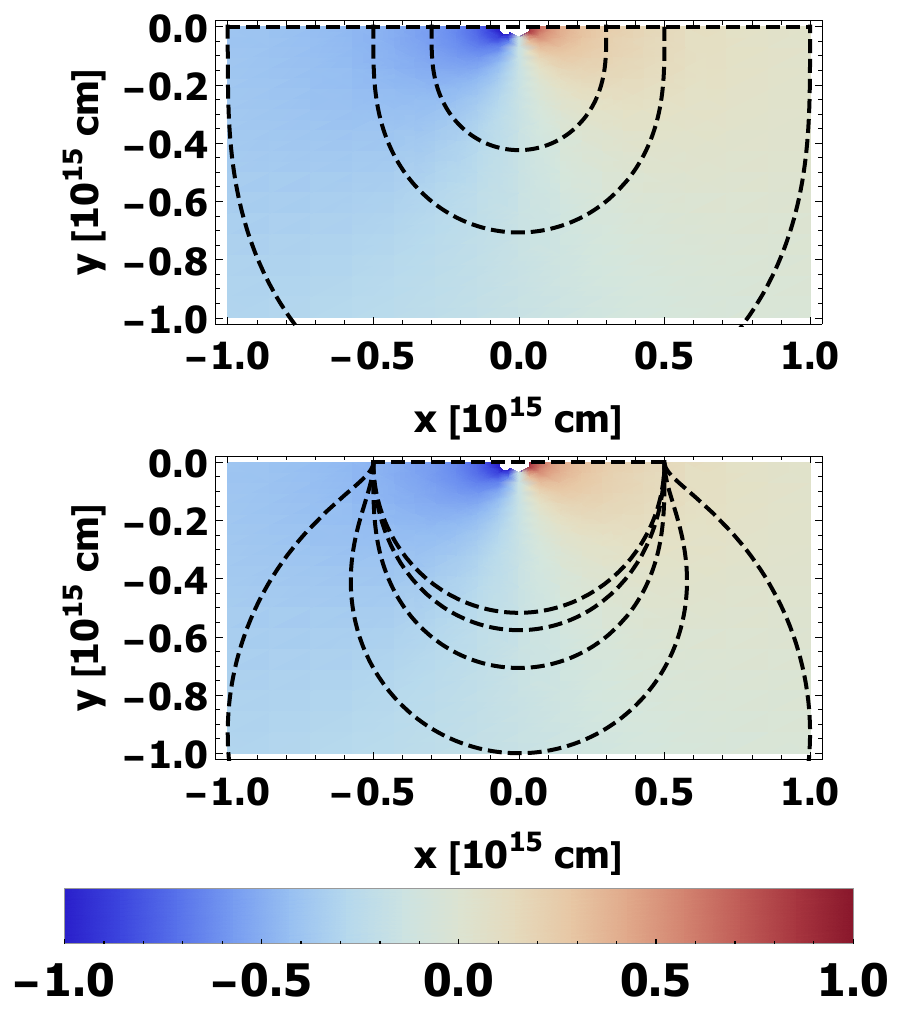}
\caption{The neutral disk as seen by the observer. Due to the Doppler shift, an observer sees a line of sight velocity profile that traces a dipole on the neutral disk (The black hole is located at $x,y=0$). The color describes the magnitude of the line of sight velocity; areas with larger (lower) redshifts are redder (bluer). The value in the color bar is normalized to $\sqrt{G M \sin i /r }$. The boundary of the disk that is illuminated by the photosphere, $R_{max}$, is plotted in dashed lines for a variety of photosphere radii, $R_{ph}$ (top) and a variety of inclinations, $i$ (bottom). The outermost $R_{max}$ in the top panel corresponds to $R_{ph}=10^{15}$ cm, and the successively smaller contours are for $R_{ph}=5 \times 10^{14}$ cm and $R_{ph}=3 \times 10^{14}$ cm. The outermost $R_{max}$ in the bottom panel corresponds to $i=15^\circ$, and the successively smaller contours corresponds to $i=30^\circ, \;45^\circ, \; 60^\circ, \; \& \; 75^\circ$. The $x$ and $y$ axes are Cartesian coordinates on the disk, in units of $10^{15}$ cm. The portion of the disk that is seen in absorption is within $0 \le \theta \le \pi $ and $R_{ph} \le r \le R_{max}$. Increasing $R_{ph}$ samples a wider region of the disk, but a larger part of the high velocity portion of the disk close to the black hole is hidden behind the photosphere. This results in an absorption profile that is deeper and thinner. A more edge-on (higher $i$) disk samples a wider region of the disk. Furthermore, the shape of the illuminated portion of a more edge-on disk covers higher velocity features, generating a deeper and wider absorption profile (see Figure \ref{changeRs}).}
\label{contour}
\end{figure}

\begin{figure}
\centering
\includegraphics[scale=0.5]{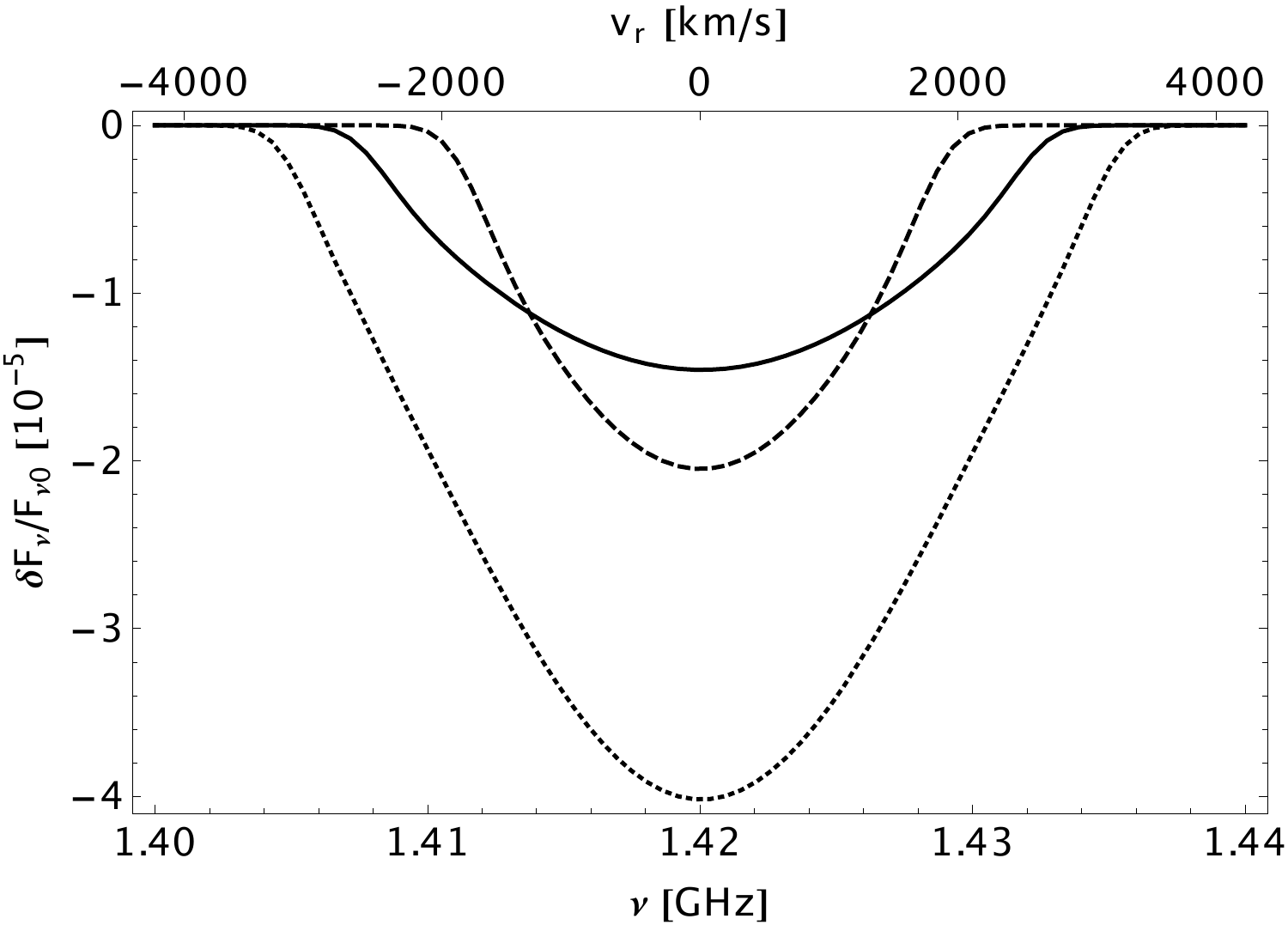}
\caption{Absorption profile of the Sgr A* photosphere for $R_{ph} = 10^{15}$ cm and an inclination angle of $i = \pi/4$ (solid), $R_{ph} = 2 \times 10^{15}$ cm and $i=\pi/4$ (dashed), as well as $R_{ph} = 10^{15}$ cm and $i=\pi/3$ (dotted). $v_r$ denotes the line-of-sight velocity. Since the disk is behind the photosphere when $r < R_{ph}$, a larger photosphere samples less of the high velocity regions of the disk, resulting in a thinner profile. However, a larger photosphere illuminates a larger disk area, resulting in more absorption. Increasing $R_{ph}$ therefore results in a deeper, but thinner profile. Edge-on geometry implies a larger portion of the disk is illuminated by the photosphere, therefore increasing the angle $i$ results in a deeper profile. A disk with higher $i$ samples higher velocity regions, resulting in a wider profile. These extra high velocity regions are sampled at the edges of the illuminated disk, in contrast to the high velocity regions sampled in systems with larger $R_{ph}$, where the extra regions are located closer to the black hole (see Figure \ref{contour}).}
\label{changeRs}
\end{figure}

%Extrapolation from Bower (2006) gives Rs = 8.2 \times 10^15 cm
Besides the multiplicative factor $\frac{N_H I_{\nu_0}}{T_S D^2}$, the problem possesses only two free parameters: the disk inclination angle, $i$, and the radius of the photosphere, $R_{ph}$. Although the intrinsic size of of Sgr A* was never measured at a wavelength as long as 21 cm, we extrapolate from the results of \cite{Bower} and find $R_{ph}$ $\sim 10^{15} \; \rm cm$. The depth of the line profile relative to the continuum, $\delta F_\nu/F_{\nu_0}$, is shown in Figure \ref{changeRs} for $N_H=10^{21} \; \rm cm^{-2}$ and $T_S = 8 \times 10^3 \; \rm K$. Note that $\delta F_\nu \propto N_H/T_S$ more generally. The distance to the Galactic center is taken as $D=8 \; \rm kpc$. A physical understanding of the absorption profiles can be obtained by looking at the line of sight velocity structure  of the disk as shown in Figure \ref{contour}. Due to the Doppler shift, an observer would detect a line of sight velocity that varies like a dipole on the disk (contours of $\sim \cos \theta / \sqrt{r}$). The portion of the disk that is illuminated by the photosphere, i.e. between $0 \le \theta \le \pi $ and $R_{ph} \le r \le R_{max}$, is seen in absorption. At every point in the illuminated disk (parameterized by the disk coordinates $r$ and $\theta$), there is a corresponding drop of flux in the frequency profile at $\nu = \sqrt{G M \sin i \cos \theta /r c^2}$ due to the absorption at that point. 

The change in the shape of the absorption profile when one changes $R_{ph}$ and $i$ can be explained by the shape of $R_{max}(R_{ph}, i)$ on the neutral disk. As Figure \ref{contour} shows, the outer bounds of the illuminated disk increases with increasing $R_{ph}$. This increases the area of the illuminated disk, generating a deeper absorption profile. However, as $R_{ph}$ increases, a larger portion of the inner disk is hidden behind the photosphere. This removes high velocity components from the integral, causing a thinner spectral profile. Indeed, Figure \ref{changeRs} demonstrates that increasing $R_{ph}$ produces a deeper and thinner absorption profile. A more edge-on (higher $i$) disk also possesses a larger illuminated disk. In this case, the inner bound is not changing, thus the wider illuminated disk also contains more high velocity portions. This generates a deeper and wider profile, as seen in Figure \ref{changeRs}.

\subsection{Disks with gaps}
\begin{figure}
\centering
\includegraphics[scale=0.5]{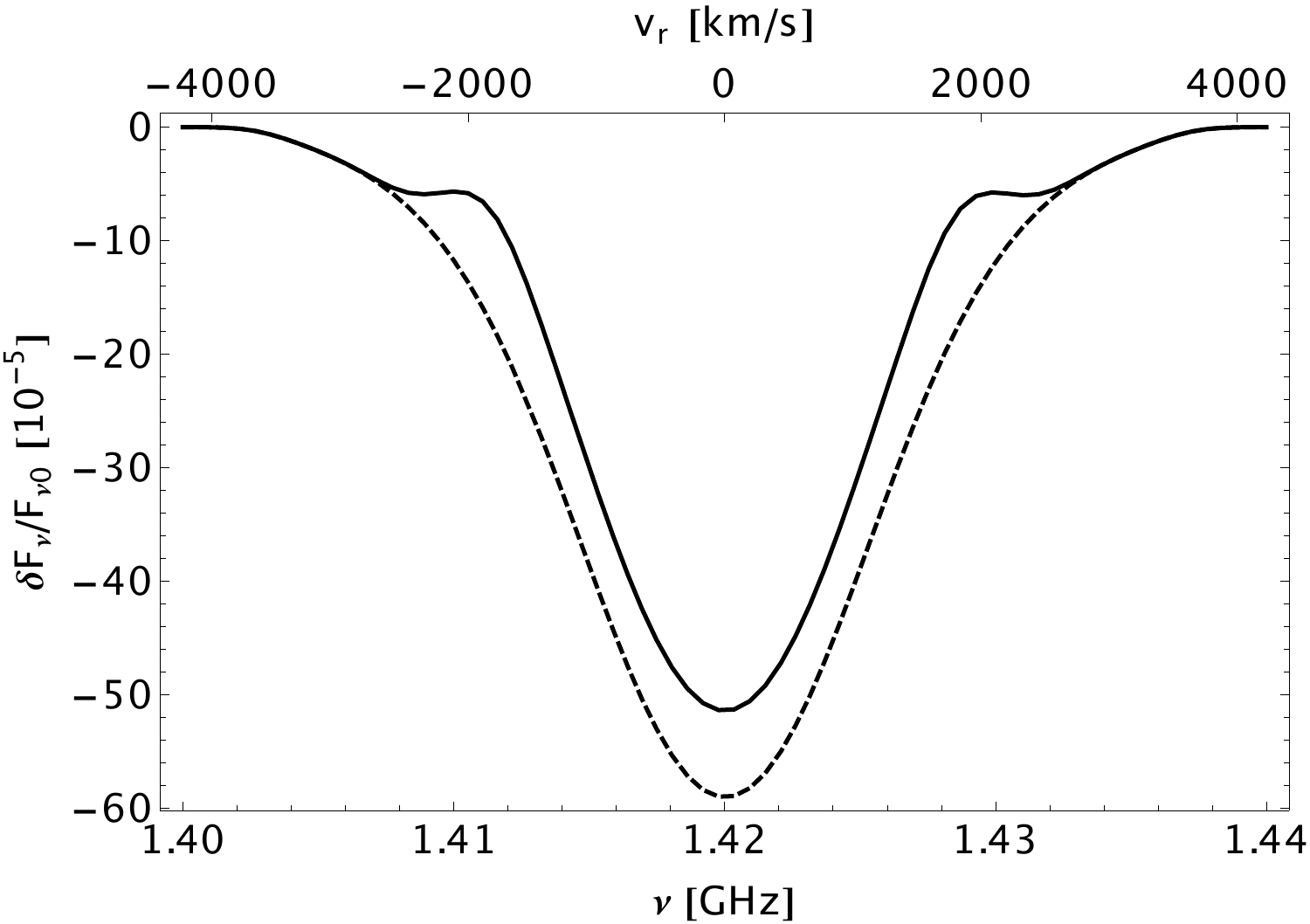}
\caption{21 cm absorption profile for a disk inclined by $80$ degrees and $R_{ph} = 2 \times 10^{15}$ cm with a gap of width $\sim R_{ph}$ (solid) overplotted against a disk with no gap (dashed). $v_r$ denotes the line-of-sight velocity.}
\label{smallgap}
\end{figure}

Using the formalism of equation (\ref{eq:form}) we can also calculate the resulting 21 cm absorption profile in the case where there are gaps in the HI disk. This is done by simply breaking the radial integral into parts,
\begin{displaymath}
   f_\nu \sim \left\{
     \begin{array}{lr}
         \int_{\theta=0}^{\theta=\pi}  \int_{r=R_{ph}}^{r=R_{max}} I \;\;\;\;\;\;\;\;\;\;\;\;\;\;\;\;\;\;\;\;\;\;\;\;\;\;\;\;\;     R_{max} < R_{in} \\
          \int_{\theta=0}^{\theta=\pi}  \int_{r=R_{ph}}^{r=R_{in}} I \;\;\;\;\;\;\;\;\;\;\;\;\;\;\;\;\;\;\;\;\;\;\;\;\;     R_{in}  < R_{max} < R_{out} \\
        \int_{\theta=0}^{\theta=\pi} \left[ \int_{r=R_{ph}}^{r=R_{in}} I + \int_{r=R_{out}}^{r=R_{max}} I \right]  \;\;\;\;\; R_{out} < R_{max} \; ,
     \end{array}
   \right.
\end{displaymath} 
where $I$ is the integrand of equation (\ref{eq:form}), $R_{in}$ is the inner radius of the gap, and $R_{out}$ is the gap's outer radius. The integral can be broken to more pieces if more gaps are present. An example profile for a disk with a gap is shown in Figure \ref{smallgap}. A disk with a gap lacks absorption on the region where the gap is located. Based on the dipolar contours of Figure \ref{contour},  the velocities corresponding to a portion between $R_{in} < r < R_{max}$ is excluded from the integral, generating a profile with visible wings. The detection of a gap can be used to constraint the presence of an intermediate mass black hole companion of Sgr A* orbiting with a semimajor axis of $\sim 10^3-10^5$ Schwarzschild radii.  

\subsection{Orbiting dense cloud}
\begin{figure}
\centering
\includegraphics[scale=0.5]{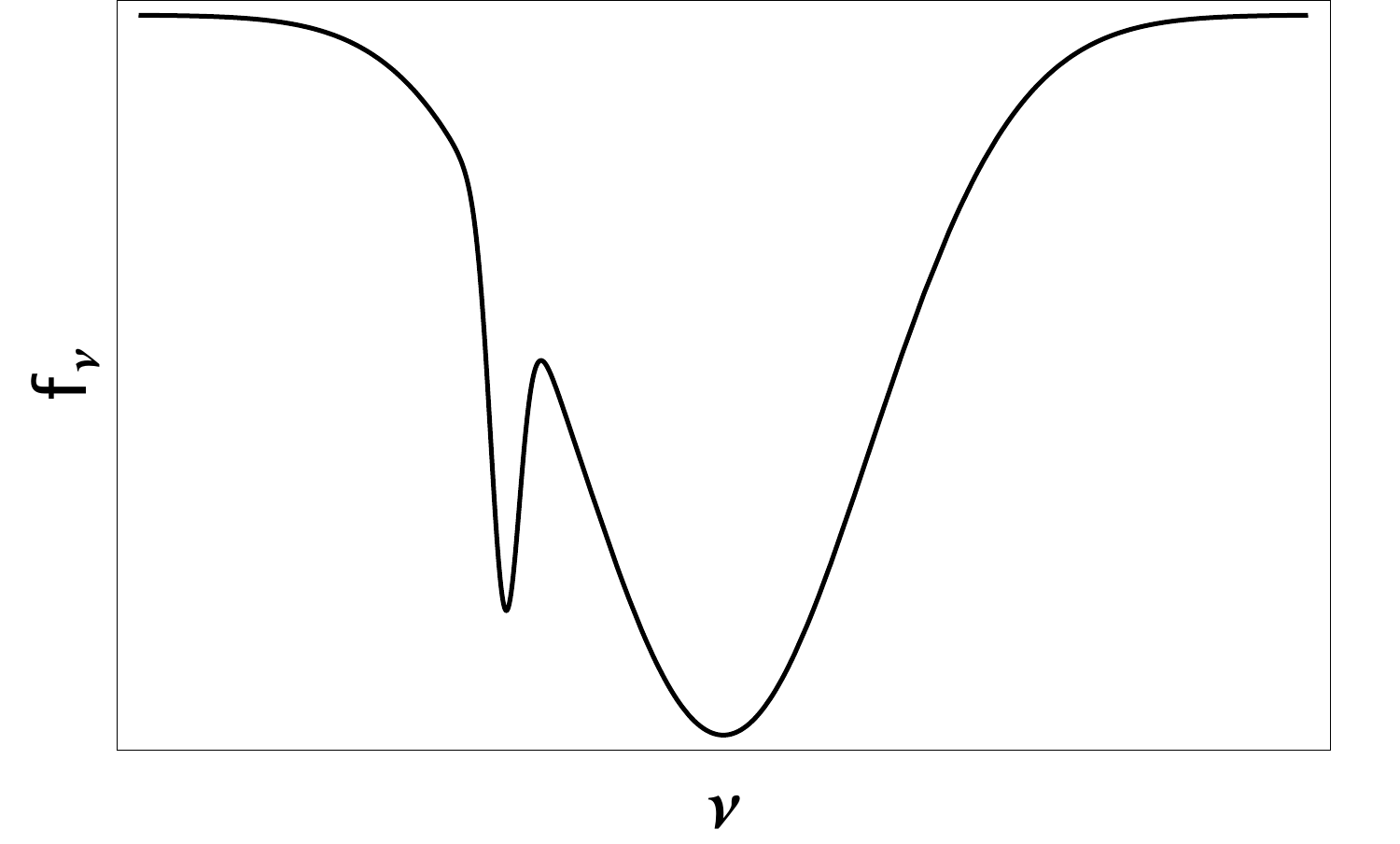}
\caption{A schematic illustration of an orbiting dense spot. As the spot orbits, it covers
different portions of the line of sight velocity structure. This resulted in extra absorption on top of
the neutral disk profile. The feature corresponding to the dense spot travels
in $\nu$ space according to $\nu_{\rm{c}}(t)$ in equation (\ref{eq:densespot}), while the HI disk profile is stationary. For
a cloud on a circular orbit, the spot absorption feature oscillates sinusoidally in frequency
space. The amplitude and frequency of the oscillation can be used to simultaneously measure
the spot orbital radius and the black hole mass up to a factor of $\sin i$. Note that a dense
spot can exist in the absence of a disk.}
\label{densespot}
\end{figure}

Another source of HI is a dense cloud that orbits in front of the 21 cm photosphere. This dense spot could be a feature in the neutral disk or a clump of self-shielded gas similar to the G2 cloud \citep{G2}. The resulting profile will be like that of Figure \ref{changeRs}, but with an extra absorption feature located where the spot resides in frequency space, as shown schematically in Figure \ref{densespot}. The core of this feature travels in $\nu$-space as the cloud orbits the black hole. If the spot orbits in a perfect circular orbit, the frequency position of the center of the feature, $\nu_c$, obeys
\beq
\nu_{\rm{c}}(t) = 1.4 \times 10^9 \rm{Hz} \left[1 - \frac{1}{c} \sqrt{\frac{G M}{r_c}} \sin i \cos \left( \sqrt{\frac{G M}{r_c^3}} (t-t_0) \right)  \right] \; ,
\eeq
where $r_c$ is the radial position of the dense spot, and $t$ the time coordinate relative to an arbitrary initial time $t_0$. The resulting effect in frequency space is a feature that oscillates sinusoidally around the 21 cm rest frame frequency. The amplitude of the oscillation is $\sqrt{\frac{GM}{c^2 r_c}} \sin i$ and the temporal frequency is $\sqrt{\frac{G M}{r_c^3}}$. Since both of these are observables, one can use them to measure simultaneously the orbital radius and the black hole mass up to a factor of $\sin i$. Plunging orbits can be treated by adding a time dependence to $r_c$,
\beq \label{eq:densespot}
\nu_{\rm{c}}(t) = 1.4 \times 10^9 \rm{Hz} \left[1 - \frac{1}{c} \sqrt{\frac{G M}{r_c(t)}} \sin i \cos \left(\sqrt{\frac{G M}{a^3}} (t-t_0) \right)  \right] \; ,
\eeq
where $a$ is the semimajor axis of the orbit and
\beq
r(t) = \frac{a(1-e^2)}{1 + e \cos \left( \sqrt{\frac{G M}{a^3}}t  \right) } \; ,
\eeq 
with $e$ being the orbital eccentricity. In addition to the inclination angle, the orbit possesses three parameters: $M$, $a$, and $e$ (or equivalently, the orbital energy, angular momentum, and the black hole mass). If the mass of the black hole is known, the oscillation amplitude and frequency can be used to determine the orbital parameters of the cloud up to the unknown inclination factor. 

\subsection{Detectability}

The signal to noise ratio of such a system is given by
\beq
\frac{S}{N} = \frac{\delta F_\nu}{\rm SEFD} \sqrt{2 t_o \Delta \nu} \; ,
\eeq 
where SEFD is the System Equivalent Flux Density of the telescope, $\Delta \nu$ the bandwidth, $t_o$ the observing time, and the factor of $\sqrt{2}$ results from the use of dual polarization observations. If Sgr A*'s intrinsic (unabsorbed) 21 cm flux density is $\sim 1 \; \rm Jy$, $N_H=10^{21} \; \rm cm^{-2}$, $T_S = 8 \times 10^{3} \; \rm K$, $i=80$ degrees, and $R_{ph} = 10^{15} \; \rm cm$, we find that $\delta F_\nu \sim 0.6 \; \rm mJy$. The Square Kilometer Array (SKA) possesses a collecting area of $10^{10} \; \rm cm^2$. A system temperature of $\sim 50$ K will then give $\rm SEFD$ $= 0.3 \; \rm Jy$. Assuming $\Delta \nu \sim 10 \; \rm MHz$, we obtain $S/N \gtrsim 10$ over 30 minutes of observations. We also note that the signal to noise ratio scales with column density, spin temperature, and observation time as
\beq
\frac{S}{N} \propto \frac{N_H}{T_S} \sqrt{t_o} \; .
\eeq 
As such, the detectability of the signal would depend on $N_H$ and $T_S$ of the observed system. In particular, for $i=80$ degrees disk, the signal to noise is given by
\beq
\frac{S}{N} \approx 9 \left[ \frac{N_H}{10^{21} \; \rm cm^{-2}} \right] \left[ \frac{8 \times 10^3 \; \rm K}{T_S} \right] \left[\frac{t_o}{1 \; \rm second}\right]^{1/2} \; .
\eeq
As we have no constraints on the values of $N_H$ and $T_S$ at the Galactic Center, the fiducial values of $N_H$ and $T_S$ considered are taken from observations of megamaser systems and theoretical considerations in our Galactic Center \citep{Loeb, YZ1}. Note that the fiducial value of $N_H$ in our calculation is modest, and column density in excess of $10^{23} \; \rm cm^{-2}$ might be possible. The spin temperature, $T_S$, is even less constrained than $N_H$; while it could be coupled to the actual gas temperature, our model considers $T_S$ as a free parameter that is not necessarily equal to the gas temperature. Our method promises to constrain the combination of the two parameters, $N_H/T_S$.

\section{Conclusions}
% Intermediate mass black hole opening up a gap in the disk

%BLURRYNESS!!!!! (INTERSTELLAR EFFECTS)
We have calculated the 21 cm absorption profile corresponding to a disk of neutral hydrogen around black holes. Spectroscopic measurements of a disk absorption profile can be used to determine the disk parameters, or to ascertain the presence of gaps in the disk even when the black hole-disk system is spatially unresolved. Furthermore, we demonstrated how an orbiting dense cloud of gas could also be detected through its time-dependent  21 cm absorption. In this case we delineated a method to determine the spot's orbital parameters from the absorption spectrum. 

%Put in VLBI or is that no relevant now?
Our methodology does not require the system to be spatially resolved. While the angular size of Sgr A*'s photosphere is $\sim 10$ milliarcseconds, interstellar blurring could render even Sgr A* unresolveable. Interstellar absorption would also deepen the absorption profile at low velocities. However, the absorption profile near the supermassive black hole is wide, and at higher velocities ($\gtrsim1000 \; \rm km/s$) the absorption should be purely from HI near the black hole. 

While we presented results for a homogeneous neutral disk with no limb darkening, our formalism is general and could be used for systems where $N_H$ or $I_{\nu0}$ are nontrivial functions of the disk coordinates, or where the line function has a significant velocity width. 

\section{Acknowledgment}
The authors would like to thank Mark Reid and Jim Moran for comments on the manuscript. This work was supported in part by NSF grant AST-1312034.


\begin{thebibliography}{99}
\bibitem[\protect\citeauthoryear{{Levin} and {Beloborodov}}{2003}]{StellarDisk} Levin Y. and Beloborodov, A. M. 2003,
apjL, 590, L33
\bibitem[\protect\citeauthoryear{Moran et al.}{2007}]{Moran} Moran, J. M. et al. 2007,
IAU Symposium, 242, 391
\bibitem[\protect\citeauthoryear{Herrnstein et al.}{2005}]{Herrnstein} Herrnstein, J. R. et al. 2005,
apj, 629, 719
\bibitem[\protect\citeauthoryear{Kuo et al.}{2011}]{Kuo} Kuo, C. Y. et al. 2011,
apj, 727, 20
\bibitem[\protect\citeauthoryear{Miyoshi et al.}{1995}]{Miyoshi} Miyoshi, M. et al. 1995,
Nature, 373, 127
\bibitem[\protect\citeauthoryear{Yusef-Zadeh and Wardle}{2012}]{YZ1} Yusef-Zadeh, F. and Wardle, M. 2012,
J. Phys.: Conf. Ser., 372, 012024
\bibitem[\protect\citeauthoryear{Yusef-Zadeh and Wardle }{2012}]{YZ2} Yusef-Zadeh, F. and Wardle, M.2012,
IAU Symposium, 287, 354
\bibitem[\protect\citeauthoryear{Loeb}{2008}]{Loeb} Loeb, A. 2008,
JCAP, 5, 8
\bibitem[\protect\citeauthoryear{Gillessen et al.}{2014}]{G2} Gillessen, S. et al. 2014,
COSPAR Meeting, 40, 992
\bibitem[\protect\citeauthoryear{Bower et al.}{2014}]{Bower} Bower, G. C. et al. 2014,
apj, 790, 1
\bibitem[\protect\citeauthoryear{Fish et al.}{2014}]{Fish} Fish, V. L. et al. 2014,
apj, 795, 134
\bibitem[\protect\citeauthoryear{Sadowski et al.}{2013}]{SG2} Sadowski, A. et al. 2013, MNRASj, 432, 478
\end{thebibliography}
\end{document}